\documentclass[secnumarabic,tightenlines,eqsecnum,floats,prd,twocolumn,superscriptaddress,showpacs,preprintnumbers,amsmath,amssymb,APS]{revtex4-1}
\usepackage{bm}
\usepackage{amsmath,amssymb,amsthm}
\usepackage{amsfonts}
\usepackage{amssymb}
\usepackage{graphicx}
\usepackage{color}
\usepackage{nicefrac}

\newcommand{\be}{\begin{equation}}
\newcommand{\ee}{\end{equation}}
\newcommand{\bea}{\begin{eqnarray}}
\newcommand{\eea}{\end{eqnarray}}

\begin{document}
\title{{\bf Adiabatic expansions for Dirac fields, renormalization, and anomalies }}

\author{J. Fernando Barbero G.}\email{fbarbero@iem.cfmac.csic.es}
\affiliation{ {\footnotesize Instituto de Estructura de la Materia, CSIC, Madrid}}
             \affiliation{ {\footnotesize  Grupo de Teor\'{\i}as de Campos y F\'{\i}sica
             Estad\'{\i}stica\\Instituto Gregorio Mill\'an, Universidad Carlos III de
             Madrid\\Unidad Asociada al Instituto de Estructura de la Materia, CSIC}}
\author{Antonio Ferreiro}\email{antonio.ferreiro@ific.uv.es}
\affiliation{ {\footnotesize Departamento de Fisica Teorica and IFIC, Centro Mixto Universidad de Valencia-CSIC.
    Facultad de Fisica, Universidad de Valencia, Burjassot-46100, Valencia, Spain}}

\author{Jose Navarro-Salas}\email{jnavarro@ific.uv.es}
\affiliation{ {\footnotesize Departamento de Fisica Teorica and IFIC, Centro Mixto Universidad de Valencia-CSIC.
    Facultad de Fisica, Universidad de Valencia, Burjassot-46100, Valencia, Spain}}

\author{Eduardo J. S. Villase\~nor}\email{ejsanche@math.uc3m.es}
\affiliation{ {\footnotesize Universidad Carlos III de Madrid, Spain}}
\affiliation{ {\footnotesize  Grupo de Teor\'{\i}as de Campos y F\'{\i}sica
             Estad\'{\i}stica\\Instituto Gregorio Mill\'an, Universidad Carlos III de
             Madrid\\Unidad Asociada al Instituto de Estructura de la Materia, CSIC}}

\date{June 20, 2018}

\begin{abstract}

We introduce an iterative method to univocally determine the adiabatic expansion of the modes of Dirac fields in spatially homogeneous external backgrounds. We overcome the ambiguities found in previous studies and use this new procedure to improve the adiabatic regularization/renormalization scheme. We provide details on the application of the method for Dirac fields living in a four-dimensional Friedmann-Lemaitre-Robertson-Walker spacetime with a Yukawa coupling to an external scalar field. We check the consistency of our proposal by working out the conformal anomaly. We also analyze a two-dimensional Dirac field in Minkowski space coupled to a homogeneous electric field and reproduce the known results on the axial anomaly. The  adiabatic expansion of the modes given here can be used to properly characterize the allowed physical states of the Dirac fields in the above external backgrounds.

\end{abstract}


\maketitle

\section{Introduction}
The renormalization of expectation values of  the stress-energy tensor---or any other relevant operator---of a quantized field in a curved space-time is more involved than in Minkowski because the curvature yields new types of divergences \cite{parker-toms, fulling, birrell-davies} (see also the recent review \cite{Hollands-WaldPR15}).  One of the earliest methods developed to deal with this issue is the so-called adiabatic regularization \cite{parker66}. It was specially designed to study the ultraviolet (UV) divergences of the stress-energy tensor of a quantized scalar field during the expansion of a homogeneous  universe \cite{parker-fulling, Bunch80}.  The renormalization subtractions in the adiabatic scheme  are integrals in the three-dimensional momentum space associated with the homogeneous hypersufaces of constant cosmic time $t$. These subtractions define a regularization procedure equivalent to the  DeWitt-Schwinger renormalization scheme \cite{Birrell, Anderson-Parker, rio1}. One  advantage of the adiabatic subtractions is that there are efficient numerical techniques to implement them \cite{Anderson}.

The adiabatic regularization and renormalization method has been generalized to deal with quantized scalar fields coupled to other types of homogeneous external fields, such as  electromagnetism \cite{Kluger91} or  classical scalars  \cite{Anderson08}. However, a systematic extension of this procedure to deal with a quantized spin-$\nicefrac{1}{2}$ field has proven to be elusive. A first substantial step in this direction was taken in a set of recent works \cite{LNT, RNT, Ghosh:2016}. In those papers several important physical examples were considered. In particular, in order to understand the reheating process after inflation, \cite{DFNT} analyzes the Yukawa coupling of a quantized Dirac field to an external scalar. Also, in \cite{FN} the coupling to an external electromagnetic field---relevant to the analysis of pair creation in strong electric fields---has been considered. The adiabatic expansion of the fermionic modes proposed in these works seemed to require a generalization of the conventional WKB ansatz used in the expansion of the scalar field modes. Although one such generalization has been proposed in those works and shown to determine appropriate subtraction terms for the renormalization of local expectation values, it was not capable of univocally fixing the proper adiabatic expansion of the field modes. The main goal of the present work is to solve this problem. The method proposed here, not related with the WKB approximation, is based on an iterative procedure involving unitary transformations. As we will see, it is robust, self-consistent and can deal with many types of external fields (gravitational, scalar, or electromagnetic). We have systematically scrutinised it and shown that its predictions for the (renormalization) subtraction terms match exactly those of the  approach discussed in \cite{LNT, RNT, DFNT, FN}. In particular we have computed the conformal and axial anomalies emerging for massless fermions and found an exact a
 greement with the results obtained by other methods.

The plan of the paper is the following. In section \ref{scalars} we provide a general and conceptual overview of the original adiabatic method for scalar fields. Section \ref{fermions} is devoted to a brief introduction of the treatment of the spin-$\nicefrac{1}{2}$ fields. Next, in section \ref{method} we present the fundamentals of the new method that we propose to uniquely determine the adiabatic expansion of fermionic modes. In section \ref{applications} we apply the method for a Dirac field in a four-dimensional Friedmann-Lema\^{i}tre-Robertson-Walker (FLRW) spacetime with a Yukawa coupling to an external scalar field. We evaluate the renormalization subtractions required to compute the conformal anomaly. We also consider two-dimensional fermions coupled to a homogeneous electric field. Finally, we conclude in section \ref{comparison} with some comments.

\section{Adiabatic expansion of scalar modes and adiabatic subtractions}\label{scalars}
The creation of particles by time-dependent  gravitational fields is a direct consequence of quantum field theory in a curved spacetime \cite{parker66, parker68}. Particle and antiparticle pairs are spontaneously produced in a way consistent with the local conservation law of the stress-energy tensor $\nabla_\mu \langle T^{\mu\nu} \rangle =0$. A prototype scenario is an isotropic expanding universe governed by the FLRW metric $ds^2 = dt^2 - a^2(t) (dx^2 + dy^2 + dz^2)$ with flat and non-compact slices of constant $t$. The time-dependence of the expansion factor  $a(t)$ is key to the particle creation process, since the creation operators at early times become a superposition of creation and annihilation operators at late times. In this scenario the isometries of the spacetime background are not enough to uniquely select a global vacuum state. We can at most determine a set of states, generically referred to as \textit{adiabatic vacua}, that play the role of the usual Minkowski vacuum. In such circumstances the early and late time vacua need not be equivalent, giving raise to particle production \cite{parker66, parker68, parker-toms, fulling, birrell-davies}. Similar arguments explain particle creation by black holes \cite{Hawking75, Waldbook, fabbri-navarro}.

To review the main ideas let us consider a scalar field $\phi$ on a FLRW background satisfying the field equation
\be (\Box + m^2 + \xi R) \phi =0 \ , \ee
and expand it in any complete orthonormal set of elementary solutions
\be \label{modedec}\phi (t,\vec{x}) = \int \mathrm{d}^3 k \, \left(a_{\vec k} f_{\vec k} (t,\vec{x}) + a_{\vec k}^{\dagger}f^*_{\vec k} (t,\vec{x})\right) \ . \ee
Moreover,  one can also  write
\be f_{\vec k}(t,\vec{x}) = \frac{1}{\sqrt{2(2\pi)^3 a^3(t)}} h_k(t) e^{i\vec k\cdot\vec x} \ , \ee
where $h_k(t)$ obeys the equation
\be \label{waveq}\ddot h_k + (\omega_k^2  + \sigma) h_k =0 \ . \ee
Here $\omega_k^2 = k^2/a^2 + m^2$ and $\sigma = (6\xi -3/4) \dot a^2/a^2 +(6\xi -3/2)\ddot a/a$. The $h_k$ are constrained to satisfy the normalization condition
\be \label{wronskian} h_k \dot h_k^* - h_k^* \dot h_k = 2i \ . \ee
It is physically reasonable to demand that the allowed solutions (defining the set of adiabatic vacua) behave as
\be \label{order0}h_k(t) \sim \frac{1}{\sqrt{\omega_k(t)}} e^{-i\int^t \omega_k(t') \mathrm{d}t'} \ , \ee
for large $k$ or, equivalently, for slow varying frequencies. To find the higher-order terms in the proposed asymptotic adiabatic series one has to enforce the normalization condition (\ref{wronskian}). This can be achieved  by introducing the WKB-type ansatz
\be \label{WKB} h_k(t) \sim \frac{1}{\sqrt{W_k(t)}} e^{-i\int^t W_k(t') \mathrm{d}t'} \ . \ee
Now, plugging this ansatz into the second-order wave equation (\ref{waveq}) the zero-order condition (\ref{order0}) determines univocally the higher-order terms in the adiabatic expansion. One finds
\be \label{W} W_k(t) \sim \omega_k + \omega_k^{(2)} + ...  \ , \ee
where
\[
\omega_k^{(2)}= \frac{1}{2} \omega_k^{-1/2} \frac{d^2}{dt^2} \omega_k^{-1/2} + \frac{1}{2} \omega_k^{-1} \sigma\,,
\]
and all the terms of odd adiabatic order are zero.

At any given order an adiabatic vacuum can be associated with each $h_k(t)$ having the asymptotic expansion (\ref{WKB}), however, in the absence of any extra condition, there is no way to choose a preferred one.  An exception to this happens for $\xi =1/6$ and $m=0$ because we can take the conformal vacuum, defined as the zero-order term in the adiabatic expansion, which is itself an exact solution.  Something similar happens in the de Sitter spacetime where a remarkable example is provided by the so-called Bunch-Davies adiabatic vacuum \cite{BD79}. This can be characterized as the unique adiabatic vacuum compatible with the ten-dimensional isometry group of de Sitter metric. For  more general cosmological scenarios see \cite {Anderson2, Ivan}.

A fundamental  advantage of the above adiabatic expansion is that it can be used to identify the UV divergences that plague the expectation values quadratic in fields. The most important observable is the stress-energy tensor $\langle T_{\mu\nu}(t,\vec{x}) \rangle$. Another relevant physical quantity is the self-correlation function  $\langle \phi^2(t,\vec{x})\rangle$, which can be used to illustrate the basic subtraction procedure \cite{parker-toms}. Assuming that the field can be written as (\ref{modedec}) one has
\be \label{phi2} \langle \phi^2(t,\vec{x})\rangle = \frac{1}{4\pi^2 a^3(t)}\int_0^\infty \mathrm{d}k\, k^2  |h_k(t)|^2 \ . \ee
If the vacuum is also assumed to be an adiabatic state one can use the expansion (\ref{WKB}) with (\ref{W}) to write a formal asymptotic expansion
\be \label{expansionphi2}\langle \phi^2\rangle \sim  \frac{1}{4\pi^2 a^3}\int_0^\infty \!\!\!\!\! \mathrm{d}k\, k^2  \left( \frac{1}{\omega_k} + \frac{1}{W_k^{(2)} }+  \frac{1}{W_k^{(4)}}+\cdots    \right)  \ee
where
\be W_k^{-1} = \omega_k^{-1} + (W_k^{-1})^{(2)} + (W_k^{-1})^{(4)} +\cdots   \ee
is the corresponding adiabatic expansion of $W_k^{-1}$ according to (\ref{W}). The expansion (\ref{expansionphi2}) tells us what the UV divergences in the $k$-integral for the expectation value (\ref{phi2}) are. These divergences are independent of the (adiabatic) vacuum state. In this case it is easy to see that the divergences appear in the $0^{\mathrm{th}}$ and $2^{\mathrm{nd}}$ order terms $(W_k^{-1})^{(0)}:= \omega_k^{-1}$ and $(W_k^{-1})^{(2)}$. The terms of higher adiabatic order ($4$ and beyond) are all finite, hence, no additional subtractions are necessary.

Once the UV divergences have been identified they can be removed by using standard renormalization methods (involving a covariant regularization scheme) and introducing appropriate covariant counterterms into the original Lagrangian. It is possible to bypass this more involved procedure by directly subtracting the relevant leading terms in the asymptotic expansion (\ref{expansionphi2}).  Moreover, to guarantee the covariance of the overall procedure one should subtract the full terms $(W_k^{-1})^{(n)}$ at the required adiabatic orders. This way one can find expressions for renormalized quantities that could differ, at most, by standard finite, covariant terms, when compared with any other manifestly covariant renormalization scheme.
Only second adiabatic order subtraction terms for $\langle \phi^2\rangle$ and four adiabatic order for $\langle T^{\mu\nu} \rangle$ are required in the minimal subtraction scheme. The renormalized expression for $\langle \phi^2\rangle_{\rm ren}$ is then defined by the momentum integral
\be \label{phi2ren} \langle \phi^2\rangle_{\rm ren} := \frac{1}{4\pi^2 a^3}\int_0^\infty \!\!\!\! \mathrm{d}k\, k^2 \left( |h_k|^2 -   \omega_k^{-1} - (W_k^{-1})^{(2)}  \right) \ee
for any expansion rate $a(t)$. During the expansion of the universe, the  quantities   $\langle \phi^2 \rangle_{\rm ren}$ and    $\langle T^{\mu\nu} \rangle_{\rm ren}$    acquire non-zero values originating in particle creation and vacuum polarization effects. We note that a natural source of finite ambiguities appears in defining the frontier between the zeroth adiabatic order and the higher adiabatic order. The zero adiabatic subtraction term can be  defined at the arbitrary scale $\mu$ instead of the natural mass scale $\mu=m$. This way one rewrites the basic equation (\ref{waveq}) as $\ddot h_k + (k^2/a^2 + \mu^2 + \tilde \sigma) h_k =0 $, where now $\tilde \sigma = (m^2 -\mu^2 + \sigma)$ is considered a quantity  of second adiabatic order.
The new value for  $\langle \phi^2\rangle_{\rm ren}(\mu)$ differs from   $\langle \phi^2\rangle_{\rm ren}(\mu=m)$ by
\bea && \langle \phi^2\rangle_{\rm ren}(m) -  \langle \phi^2\rangle_{\rm ren}(\mu) =\\&&  \frac{1}{16 \pi^2}\left( m^2-\mu^2+m^2\log\frac{m^2}{\mu^2} - (\xi- \frac{1}{6}) \frac{R}{3} \log\frac{m^2}{\mu^2}\right). \nonumber  \eea
Note that this agrees with the results in \cite{Waldbook, Hollands-WaldPR15, Hollands-WaldCMP}. All renormalization methods give equal vacuum expectation values in arbitrary (globally hyperbolic) curved spacetimes up to finite, covariant ambiguities of the form found above.

We want to remark that the above adiabatic subtraction procedure satisfies all the conditions necessary to have a consistent renormalization procedure in a curved spacetime (Wald axioms) \cite{Waldbook, Hollands-WaldPR15, Hollands-WaldCMP}:
\begin{enumerate}
\item The subtractions are state-independent.
\item The subtractions are local. This is so because the relevant subtraction terms $(W_k^{-1})^{(n)}$ do not involve time integrals and depend locally on the expansion rate $a(t)$. The time integral in the phase term of (\ref{WKB}) does not play any role in the asymptotic expansion of physical expectation values such as  (\ref{expansionphi2}).
\item Covariance is preserved, in particular  $\nabla_\mu \langle T^{\mu\nu} \rangle_{\rm ren} =0$. This is so because the concept of ``adiabaticity'' is consistent with  general covariance. In particular one can check that  $\nabla^\mu  T_{\mu\nu}^{(n)} =0$, at any adiabatic order $n$.
\end{enumerate}
In addition, in Minkowski space the adiabatic subtractions are equivalent to the conventional normal ordering prescription for $\mu=m$ and, hence,
  $ \langle T^{\mu\nu} \rangle_{\rm ren} =0 = \langle \phi^2 \rangle_{\rm ren}$. Throughout this work we maintain the conventional assumption $\mu=m$ in all the computations.

One can also be concerned about the possibility of oversubtraction. For instance, if one (incorrectly) subtracts terms up to and including the sixth adiabatic order for the renormalization of the stress-energy tensor, one finds spurious covariant terms of the form $m^{-2} t_{\mu\nu}$, where $t_{\mu\nu}$ is a geometric conserved tensor of  mass dimension six made out of local curvature tensors and covariant derivatives. These terms are not physically allowed since they are not continuous in the massless limit. This reinforces the choice  of the minimal subtraction scheme mentioned above.

A consequence of the subtractions needed to obtain well-defined physical quantities is that field relations that are valid in the classical theory cannot be fully respected at the quantum level. For instance, for $\xi=1/6$ and $m=0$, the scalar field obeys the conformally invariant field equation
\be (\Box + \frac{1}{6} R) \phi =0 \ . \ee
However, when evaluating the expectation value
\be \langle \phi(\Box + \frac{1}{6} R) \phi \rangle_{\rm ren} \ , \ee
one gets a surprising result: instead of zero, as naively expected from the classical theory, one finds
\be \langle \phi(\Box + \frac{1}{6} R) \phi \rangle_{\rm ren}=  \frac{-1}{2880 \pi^2} [ \Box R - (R^{\mu\nu} R_{\mu\nu} - \frac{1}{3} R^2)]\ , \ee
which in fact turns out to be the negative of the trace anomaly of the massless conformal field in a FLRW spacetime. We note in passing that in this situation there is no particle production and the bulk of the result can only be  attributed to vacuum polarization effects.

Finally, a byproduct of the adiabatic expansion is the fact that one can naturally demand that any physical vacuum has to be UV regular \cite{Ivan} in the sense that the large $k$ behavior of modes $h_k(t)$ defining the state must approach the one found in Minkowski space at the appropriate rate. This condition is closely related to the Hadamard condition in generic backgrounds \cite{Pirk93, Hollands01, Fewster-Verch13}. For scalar fields, and in order to guarantee that  the renormalized value of $\langle T^{\mu\nu} \rangle$ is finite, the large $k$ behavior of the modes must be fixed as follows
\bea
|h_k| &\sim& \frac{1}{\sqrt{\omega_k + \omega^{(2)}_k+ \omega^{(4)}_k}} + \mathcal{O}(k^{-11/2}) \nonumber \\
&\sim& \frac{1}{(k/a)^{1/2}} -\frac{C(a,m,\xi)}{4(k/a)^{5/2}} +\frac{D(a,m,\xi)}{32(k/a)^{9/2}}+ \mathcal{O}(k^{-11/2})\nonumber
\eea
with
\begin{align}
C(a,m,\xi)&:= \xi' R +m^2\,, \nonumber\\
D(a,m,\xi)&:=4 m^2\frac{\dot a^2}{a^2}+5 m^4
+ 10m^2R-4 \xi' R \frac{
   \dot a^2}{a^2}\nonumber \\
  &+24 \xi' \frac{\dddot a}{a}\frac{\dot a}{a}+5\xi'^2
   R^2 +2\xi' \Box R\nonumber \\
  &+10\xi' m^2 R+\frac{8}{3}\xi' R^2-6\xi' R_{\mu\nu}R^{\mu \nu}\,,\nonumber
\end{align}
and $\xi':=\xi -\frac{1}{6}$
and, similarly, for $ |\dot h_k(t)|$. One denotes  such  vacua as  fourth order adiabatic states. One can check, for instance, that the Bunch-Davies vacuum  is indeed of  adiabatic order $n=\infty$. \\

\section{Quantized Dirac fields coupled to homogeneous external fields}\label{fermions}
\subsection{Dirac fields in an homogeneous expanding universe}

The covariant Dirac equation in curved spacetime is given by \cite{parker-toms, birrell-davies}
\be \label{Dirac0}i{\bm{\gamma}}^{\mu}\nabla_{\mu} \psi - m\psi =0
\ee
where ${\bm{\gamma}}^{\mu}(x) $ are the spacetime-dependent Dirac-matrices satisfying the condition $\{{\bm{\gamma}}^{\mu}, {\bm{\gamma}}^{\nu}\}= 2g^{\mu \nu}$ and  $\nabla_{\mu}$ is the covariant derivative associated with the  spin connection.
In a  spatially flat FLRW universe, the matrices $\bm{\gamma}^{\mu}(t)$ are related to  the constant Dirac matrices in Minkowski spacetime $\gamma^{\alpha}$, obeying  $\{{\gamma}^{\alpha}, {\gamma}^{\beta}\}= 2\eta^{\alpha \beta}$, by the simple relations
\be \bm{\gamma}^0(t) = \gamma^0 \ ;  \  \ \ \ \ \ \ \   \bm{\gamma}^i (t) = \gamma^i/a(t) \ . \ee
The Dirac equation takes then the form
\be \left(i\gamma^0\partial_0+ \frac{3i}{2}\frac{\dot a }{a}\gamma^0  +\frac{i}{a}\vec{\gamma}\cdot\vec{\nabla} -m\right)\psi=0 \ . \label{diracgamma} \ee
Introducing the expansion
\[
\psi(t,\vec{x})=\int \mathrm{d}^3k\,\,  \psi_{\vec{k}}(t)e^{i\vec{k}\cdot\vec{x}}
\]
and using the Dirac representation for the $\gamma^\alpha$ matrices
\be
\gamma^0 =
\left( {\begin{array}{rr}
 I & 0  \\
 0 & -I  \\
 \end{array} } \right)\,,
\hspace{1cm} \vec\gamma = \left( {\begin{array}{rr}
 0 & \vec\sigma  \\
 -\vec\sigma & 0  \\
 \end{array} } \right)
 \ee
 where $\vec{\sigma}$ are the usual Pauli matrices, we can write \cite{LNT, RNT}
\bea \label{psik}
&\psi_{\vec{k}}(t)=
\left( {\begin{array}{c}
  \frac{1}{\sqrt{(2\pi)^3 a^3(t)}}h^I_{{k}}(t) \xi_{\lambda} (\vec{k}) \\
  \frac{1}{\sqrt{(2\pi)^3 a^3(t)}}h^{II}_{{k}}(t)\frac{\vec{\sigma}\cdot \vec{k}}{k} \xi_{\lambda} (\vec{k})\\
 \end{array} } \right)
\eea
where $k:=|\vec k|$ and $ \xi_{\lambda} (\vec{k})$ is a constant  normalized two-component spinor $\xi_{\lambda}^{\dagger}\xi_{\lambda}=1$ such that
 $\frac{\vec{\sigma}\cdot\vec{k}}{2k}\xi_{\lambda}= \lambda \xi_{\lambda}$.
$\lambda ={\pm} 1/2$  represents the eigenvalue for the helicity, or spin component along the $\vec{k}$ direction.
$h_{{k}}^I$ and $h_{{k}}^{II}$ are scalar functions,  which as a consequence of (\ref{diracgamma}), obey the coupled first-order equations
\begin{align}
\label{eqhI,II} i \partial_t h_{{k}}^{I}&=m h_{{k}}^{I}+ \frac{k}{a} h_{{k}}^{II}\,,\\
  i \partial_t h_{{k}}^{II}&=\frac{k}{a} h_{{k}}^{I}-m h_{{k}}^{II}   \label{eq:hII} .
\end{align}
The normalization condition for the four-spinor is
\bea \label{normalization} &|h_{{k}}^I(t)|^2 +   |h_{{k}}^{II}(t)|^2 =1 \ . \eea
This condition enforces the standard anticommutator relations for creation and annihilation operators  defined by the expansion
\be \psi(t, \vec{x})= \int \mathrm{d}^3k\sum_{\lambda = {\pm 1/2}} ( B_{\vec{k}, \lambda } u_{\vec{k}, \lambda}(t, \vec{x}) +  D^{\dagger}_{\vec{k}, \lambda} v_{\vec{k}, \lambda} (t, \vec{x}) )\ , \ee
where
\be u_{\vec{k}, \lambda} (t,\vec{x}) = \frac{1}{\sqrt{(2\pi)^{3}a^3(t)}} e^{i\vec{k}\cdot\vec{x}}\left( {\begin{array}{c}
 h^I_{{k}}(t)  \xi_{\lambda}\\
h^{II}_{{k}}(t) \frac{\vec{\sigma}\cdot\vec{k}}{k}\xi_{\lambda}\\
 \end{array} } \right) \ . \ee
These modes are normalized with respect to the Dirac scalar product
\be (u_{\vec{k}, \lambda}, u_{\vec{k'}, \lambda'})= \int \mathrm{d}^3x \, a^3{u^{\dagger}}_{\vec{k}, d\lambda}u_{\vec{k'}, \lambda'}= \delta (\vec{k}- \vec{k'}) \delta_{\lambda \lambda'}
\ee

The orthogonal modes $ v_{\vec{k}, \lambda} (t, \vec{x})$ are obtained by charge conjugation
\[
v_{\vec{k}, \lambda}=Cu_{\vec{k}, \lambda}=-i\gamma^2u^*_{\vec{k}, \lambda}.
\]
We then have
\bea \{B_{\vec{k}, \lambda}, B_{\vec{k}', \lambda'}^{\dagger}\}= \delta^3 (\vec{k}-\vec{k}')\delta_{\lambda \lambda'} \\  \{B_{\vec{k}, \lambda}, B_{\vec{k}', \lambda'}\}= 0=  \{B_{\vec{k}, \lambda}^{\dagger}, B_{\vec{k}', \lambda'}^{\dagger}\} \ , \eea
and similarly for the  $D_{\vec{k}, \lambda}$, $D_{\vec{k}, \lambda}^{\dagger}$ operators.

We can add to the Dirac equation (\ref{Dirac0}) a Yukawa coupling with an external scalar field $\Phi$
\be \label{Dirac1}(i{\gamma}^{\mu}\nabla_{\mu}  - m -g_Y\Phi)\psi =0 \ . \ee
If the classical field $\Phi$ is also assumed to be homogeneous $\Phi= \Phi (t)$, as the classical background geometry, the modes for the quantized Dirac field can be expressed as above. However, the basic equations (\ref{eqhI,II}) are now replaced by \cite{DFNT}
\begin{align}
i\partial_th_{{k}}^I &=(m+ s(t))h_{{k}}^I+\frac{k}{a(t)}h_{{k}}^{II}\label{eqhI}\\
i\partial_th_{{k}}^{II} &=\frac{k}{a(t)}h_{{k}}^{I}-(m+ s(t))h_{{k}}^{II}\, , \label{IIbis}
\end{align}
where $s(t) := g_Y\Phi(t)$.

 \subsection{Two-dimensional Dirac fields in an homogeneous electric field}

The Dirac equation in the presence of an external homogeneous electric field and a background metric of the form $ds^2 = dt^2 -a^2(t)dx^2$ is
\[(i {\gamma}^{\mu}\nabla_{\mu}-m)\psi=0,
\]
where now
\[
\nabla_{\mu} := \partial_{\mu}-\Gamma_{\mu} -i q A_{\mu}\,,
\]
$\Gamma_\mu$ is the spin connection and $A_\mu$ is the vector potential. We assume  $A_\mu=(0, -A(t))$. It is also useful to use here the Weyl representation for the Dirac matrices in flat Minkowski space
\bea
\gamma^0 =
\left( {\begin{array}{rr}
 0 & 1  \\
 1& 0  \\
 \end{array} } \right),\hspace{1cm}
\gamma^1 =  \left( {\begin{array}{rr}
 0 & 1  \\
 -1& 0  \\
 \end{array} } \right)
 \ . \eea
Expanding the field in  momentum modes
\be \psi(t, x)=\int_{-\infty}^{+\infty} \mathrm{d}k\,\, \psi_{k}(t)e^{ikx} \ , \ee the Dirac equation takes the form \cite{FN}
\bea
\left(\partial_0+\frac{\dot{a}}{2a}+\frac{i}{a}(k+q A)\gamma^0\gamma^1 +im\gamma^0\right)\psi_k=0 \ .
\eea
We can build two independent spinor solutions
\bea
u_{k}(t, x)&=&\frac{e^{ikx}}{\sqrt{2\pi a(t)}}  \left( {\begin{array}{c}
h^{I}_k(t)   \\
-h^{II}_k (t) \\
\end{array} }\right) \\
v_{k}(t, x)&=&\frac{e^{-ikx}}{\sqrt{2\pi a(t)}}  \left( {\begin{array}{c}
h^{II*}_{-k} (t)  \\
h^{I*}_{-k}(t)  \\
\end{array} } \right)
\ , \eea
where $ h^{I}_k(t)$ and $ h^{II}_k(t)$ must satisfy the equations
\begin{align} \label{system}
i\partial_t h^{I}_k&=-\frac{1}{a}(k+q A)h^{I}_k-m h^{II}_k\,,\\
i\partial_t h^{II}_k&=-m h^{I}_k+\frac{1}{a}(k+q A)h^{II}_k\,,
\end{align}
and also obey the normalization condition \eqref{normalization}.

\section{An iterative approach to the adiabatic expansion for Dirac fields}\label{method}

As mentioned in the previous section the dynamical content of the modes of the Dirac fields can be encoded as a pair of complex functions of time $h_k^I(t)$ and $h_k^{II}(t)$ satisfying, for all values of $t$, the normalization condition \eqref{normalization}. For all the models considered in the previous section these functions satisfy a system of ODE's of the form
\begin{equation}\label{basic_equation}
  i\partial_t \left(\begin{array}{c}
                      h^I \\
                      h^{II}
                    \end{array}\right)=
                    \left(\begin{array}{rr}
                    \alpha(t) & \beta(t) \\
                    \beta(t) &  -\alpha(t)
                    \end{array}\right)
                    \left(\begin{array}{c}
                      h^I \\
                      h^{II}
                    \end{array}\right)\,.
\end{equation}
This is, in fact, a Schr\"{o}dinger equation
\begin{equation}\label{basic_equation_condensed}
  i\partial_t \bm{h} = \bm{H}(t) \bm{h}
\end{equation}
 defined by a matrix, time-dependent Hamiltonian
\begin{equation}\label{Hamiltonian}
\bm{H}(t)=\left(\begin{array}{rr}
                    \alpha(t) & \beta(t) \\
                    \beta(t) &  -\alpha(t)
                    \end{array}\right)
                  \,,
\end{equation}
acting on a two-component wave function $\bm{h}$. As a consequence of this, the normalization condition \eqref{normalization} is automatically satisfied for all times for initial data satisfying $|h^I(t_0)|^2+|h^{II}(t_0)|^2=1$.

For generic choices of the functions $\alpha(t)$ and $\beta(t)$ this Schr\"{o}dinger equation cannot be solved in closed form. In the following we develop an iterative procedure to transform \eqref{basic_equation} in such a way that we get consistent approximations for its solutions at the desired adiabatic order. The three basic features of the proposed procedure are:

\begin{enumerate}
\item At each step we introduce a change of variables defined by a unitary transformation. This guarantees that the normalization condition \eqref{normalization} is automatically preserved.
\item At each step it will be evident how to truncate the resulting equations to reach the desired order in the adiabatic approximation.
\item The truncated equations involve a diagonal time-dependent Hamiltonian and, hence, can be trivially solved.
\end{enumerate}

The starting point is to diagonalize the matrix Hamiltonian
\[
\bm{H}(t)=\bm{U}_0(t)\bm{D}_0(t)\bm{U}_0^\dagger (t)
\]
with
\begin{equation}\label{D_0}
\bm{D}_0(t)=\left(\begin{array}{cc}
\omega_0(t)&0\\
\quad&\quad\\
0&-\omega_0(t)
\end{array}\right)\,,
\end{equation}
\begin{equation}\label{U_0}
 \bm{U}_0(t)=\left(\begin{array}{cc}
\sqrt{\frac{\omega_0(t)+\alpha(t)}{2\omega_0(t)}} & {\scriptstyle{\sigma_\beta}} \sqrt{\frac{\omega_0(t)-\alpha(t)}{2\omega_0(t)}} \\
\quad&\quad\\
{\scriptstyle{\sigma_\beta}}\sqrt{\frac{\omega_0(t)-\alpha(t)}{2\omega_0(t)}}  & -\sqrt{\frac{\omega_0(t)+\alpha(t)}{2\omega_0(t)}}
\end{array}\right)\,,
\end{equation}
where $\omega_0(t):=\sqrt{\alpha^2(t)+\beta^2(t)}$ and $\sigma_\beta$ denotes the sign of $\beta(t)$ (which, in all the cases of interest, will be well defined as $\beta(t)$ will never vanish).

We introduce now another set of variables
\[
\bm{h}_0(t):=\bm{U}^\dagger_0(t)\bm{h}(t).
\]
They satisfy the new Schr\"{o}dinger equation
\begin{equation}\label{equation_0}
   i\partial_t \bm{h}_0 = \bm{H}_0(t) \bm{h}_0\,,
\end{equation}
where
\[
\bm{H}_0(t):=\bm{D}_0(t)-i \bm{U}_0^\dagger(t)\partial_t \bm{U}_0(t)
\]
has the explicit form
\begin{equation*}
  \bm{H}_0=\left(\begin{array}{cc}
  \omega_0 & i \scriptstyle{\sigma_\beta}\displaystyle\frac{\omega_0\dot{\alpha}-\alpha\dot{\omega}_0}{2\omega_0\sqrt{\omega_0^2-\alpha^2}} \\
  \quad&\quad\\
 -i\scriptstyle{\sigma_\beta}\displaystyle\frac{\omega_0\dot{\alpha}-\alpha\dot{\omega}_0}{2\omega_0\sqrt{\omega_0^2-\alpha^2}} & -\omega_0
  \end{array}\right)\,.
\end{equation*}
Here we have used dots to represent time derivatives and lightened the notation by not writing the explicit time dependence.

The key observation at this point is to realize that the lowest adiabatic order of the off-diagonal terms of $\bm{H}_0$ is one unit higher than that of the corresponding ones in $\bm{H}$. If we repeat now the previous procedure (diagonalization of the Hamiltonian and ``unitary change of variables'') this same behavior will occur at each iteration order. Once the non-diagonal elements of the Hamiltonian surpass a certain adiabatic order $n$ we will discard them. By doing this the resulting Schr\"{o}dinger equation can then trivially solved (because the corresponding Hamiltonian is diagonal) and, by undoing the sequence of changes of variables arrive at an approximate solution to \eqref{basic_equation}.

If at a certain iteration order $j\geq0$ we have $\bm{h}_j$ obtained from the Schr\"{o}dinger equation associated with the Hamiltonian $\bm{H}_j$ the objects in the $j+1$ step are given by
\begin{align}\label{step_k}
  \bm{h}_{j+1} & =\bm{U}^\dagger_{j+1}\bm{h}_j\,, \\
  \bm{H}_{j+1}=&\bm{D}_{j+1}-i\bm{U}^\dagger_{j+1}\partial_t \bm{U}_{j+1}\,,
\end{align}
with the diagonal matrix $\bm{D}_{j+1}$ and the unitary  matrix  $\bm{U}_{j+1}$ obtained by diagonalizing $\bm{H}_j$:
\begin{align}\label{step_k_1}
  \bm{H}_j & =\bm{U}_{j+1}\bm{D}_{j+1}\bm{U}^\dagger_{j+1}\,.
\end{align}
Notice that $ \bm{h}_{j+1}$ satisfies the Schr\"{o}dinger equation
\[
i\partial_t\bm{h}_{j+1}=\bm{H}_{j+1}\bm{h}_{j+1}\,.
\]
The explicit expressions for the $\bm{H}_j$, $\bm{U}_j$ and $\bm{D}_j$ are
\begin{align*}
  \bm{H}_j=&\left(
  \begin{array}{cc}
    \omega_j &  S_{j-1}\frac{\omega_{j-1} \dot{\omega}_j-\omega_j\dot{\omega}_{j-1}}{2\omega_j\sqrt{\omega_j^2-\omega_{j-1}^2}} \\
    S^*_{j-1}\frac{\omega_{j-1} \dot{\omega}_j-\omega_j\dot{\omega}_{j-1}}{2\omega_j\sqrt{\omega_j^2-\omega_{j-1}^2}} & - \omega_j
  \end{array}\right)\\
  \bm{U}_j=&\left(
  \begin{array}{cc}
   \sqrt{\frac{\omega_j+\omega_{j-1}}{2\omega_j}}\quad & iS_{j-1} \sqrt{\frac{\omega_j-\omega_{j-1}}{2\omega_j}} \\
    iS_{j-1} \sqrt{\frac{\omega_j-\omega_{j-1}}{2\omega_j}}\quad   &(-1)^{j+1} \sqrt{\frac{\omega_j+\omega_{j-1}}{2\omega_j}}
  \end{array}\right)\\
 \bm{D}_j=&\left(\begin{array}{cc}
\omega_j&0\\
\quad&\quad\\
0&-\omega_j
\end{array}\right)
\end{align*}
where the positive frequencies $\omega_j$ satisfy the recurrence
\begin{equation}\label{omegas}
\omega_{j+1}^2=\omega_j^2+\frac{(\omega_{j-1}\dot{\omega}_j-\omega_j\dot{\omega}_{j-1})^2}{(2\omega_j)^2(\omega_j^2-\omega_{j-1}^2)}\,,
\end{equation}
with initial data
\[
\omega^2_0=\alpha^2+\beta^2\,,\quad \omega^2_1=\omega^2_0+\frac{(\omega_0\dot{\alpha}-\dot{\omega}_0\alpha)^2}{4\omega_0^2\beta^2}\,.
\]
The $S_j$ coefficients are given by
\begin{align}
S_j=&\left\{\begin{array}{rl}
             -i s_{j-1}\,,\quad & j\,\,\rm{even} \\
             s_{j-1}\,,\quad & j\,\,\rm{odd}
           \end{array}
\right.
\end{align}
where
\[
s_j=s_{j-1}\mathrm{sign}(\omega_{j-1}\dot{\omega}_j-\dot{\omega}_{j-1}\omega_j)\,,\quad j\geq 1\,,
\]
and
\[
s_0=\sigma_\beta\mathrm{sign}(\alpha\dot{\omega}_0-\omega_0\dot{\alpha})\,.
\]

Several comments are in order now:

\medskip

First it is important to notice that the lowest adiabatic weight of the non-diagonal terms of the Hamiltonian $\bm{H}_j$ is larger than $j$ (although even higher orders may be present). This fact suggests a terminating criterion to obtain an approximate solution valid at adiabatic order $n$: replace the Hamiltonian $\bm{H}_n$ by its diagonal part $\bm{D}_n$ and approximate $\bm{h}_n$ by $\tilde{\bm{h}}_n$ satisfying the Schr\"{o}dinger equation
\[
i\partial_t \tilde{\bm{h}}_n=\bm{D}_n\tilde{\bm{h}}_n\,.
\]
This way we get
\begin{align*}
\tilde{\bm{h}}_n(t)&=\widetilde{\bm{U}}_n(t,t_0) \mathfrak{h}(t_0)
\\
&:=\left(\begin{array}{cc}
                             \exp(-i\int^t_{t_0} \omega_n) & 0 \\
                             0 & \exp(i\int^t_{t_0} \omega_n)
                           \end{array}\right)\left(\begin{array}{c}
                                                     1 \\
                                                     0
                                                   \end{array}\right)\,,
\end{align*}
where our choice of initial data selects positive frequencies.

The final form for the approximate solution $\bm{h}(t|n)$  of \eqref{basic_equation} can be obtained by undoing the unitary transformations introduced above
\begin{equation}\label{sol_approx}
  \bm{h}(t)\sim\bm{h}(t|n):=\bm{U}_0(t)\bm{U}_1(t)\cdots\bm{U}_n(t)\widetilde{\bm{U}}_n(t,t_0)\mathfrak{h}(t_0)\,.
\end{equation}
From this last expression it is straightforward to obtain the adiabatic expansion of $\bm{h}$ to order $n$. In the next section we will apply the procedure that we have just described to the computation of different types of anomalies for systems involving quantized Dirac fermions in an expanding universe. In the first case the Dirac field will be coupled to a homogeneous background scalar with a Yukawa interaction term \cite{rio1} and in the second to an external homogeneous electric field and background metric of the FLRW type in $1+1$ dimensions \cite{FN}.

\section{Applications}\label{applications}

We apply now the method developed in the previous section to some anomaly computations in two important physical examples.

\subsection{Dirac fermion coupled to a homogeneous background scalar field with a Yukawa interaction in an expanding universe}

By comparing equations \eqref{eqhI} and \eqref{IIbis} with \eqref{Hamiltonian} we see that
\begin{align}\label{equations_Yukawa}
  \alpha(t) & =m+s(t)\,, \\
  \beta(t)  &=\frac{k}{a(t)}\,,
\end{align}
where $s(t)$ describes the homogeneous scalar field coupled to the Dirac field, $a(t)$ is the scale factor, $m$ is the Dirac field mass and $k$ labels the Fourier modes (we will drop the $k$ indices that should appear in $h^I$ and $h^{II}$). The adiabatic order of $s$ is one and the scale factor $a$ has adiabatic order zero.

In order to work out the conformal anomaly (see \cite{DFNT} and later on), we need to compute $\bm{h}(t|4)$ by using \eqref{sol_approx}. It is important to keep in mind that the relevant part of $\widetilde{\bm{U}}_4(t,t_0)\mathfrak{h}(t_0)$ is just the phase $\exp\left(-i\int^t \omega_4\right)$ which drops out of the computations of the anomaly, which is closely related to $\langle\overline{\psi}\psi\rangle$. For this reason we will not give the explicit form of $\omega_4$. A similar cancelation of the potential contribution of the non-local term $\int^t \omega_4$ happens in all components of the vacuum expectation values of the renormalized stress-energy tensor. Only local terms like $\omega_4(t) $  are relevant to determine the renormalization subtraction terms at the instant $t$, in accordance with the locality requirement of any renormalization scheme.

The product $\bm{U}_0(t)\bm{U}_1(t)\cdots\bm{U}_4(t)$ can be exactly computed in principle, however we only need its adiabatic expansion to fourth order. After a long but conceptually direct computation (that can be conveniently performed with the help of Mathematica\texttrademark) we get
\[
h^I\sim\sqrt{\frac{\omega+m}{2\omega}}\Big(1+(\omega-m)\sum_{n=1}^4 \phi^{(n)} \Big)\exp\left(-i\int \omega_4\right)
\]
where $\omega^2=m^2+k^2/a^2$ and
\begin{align*}
  \phi^{(1)} =&-\frac{i m \dot{a}}{4 a \omega ^3}+\frac{s}{2 \omega ^2}\hspace*{5.4cm}
\end{align*}
\begin{align*}
   \phi^{(2)} =&
   -\frac{m \ddot{a}}{8 a \omega ^4}+\frac{11 m^3 \dot{a}^2}{32 a^2 \omega^6}-\frac{m^2 \dot{a}^2}{32 a^2 \omega^5}-\frac{m \dot{a}^2}{8 a^2 \omega ^4}\\
   &+\frac{7 i m^2 s \dot{a}}{8 a \omega^5}+\frac{i m s \dot{a}}{8 a \omega ^4}-\frac{i s \dot{a}}{4 a\omega^3}-\frac{5 m s^2}{8 \omega^4}-\frac{s^2}{8 \omega^3}
   -\frac{i \dot{s}}{4 \omega^3}
\end{align*}
 \begin{align*}
  \phi^{(3)} =&+\frac{i \dddot{a} m}{16 a \omega ^5}+\frac{129 i m^5
   \dot{a}^3}{128 a^3 \omega ^9}-\frac{3 i m^4
   \dot{a}^3}{32 a^3 \omega ^8}-\frac{19 i m^3 \dot{a}
   \ddot{a}}{32 a^2 \omega ^7}\\&-\frac{97 i m^3
   \dot{a}^3}{128 a^3 \omega ^7}+\frac{i m^2
   \dot{a}^3}{32 a^3 \omega ^6}+\frac{i m
   \dot{a}^3}{16 a^3 \omega ^5}+\frac{i m^2 \dot{a}
   \ddot{a}}{32 a^2 \omega ^6}+\frac{i m \dot{a} \ddot{a}}{4 a^2 \omega
   ^5}\\&+\frac{9 m^2 s \ddot{a}}{16 a
   \omega ^6}+\frac{m s \ddot{a}}{16 a \omega ^5}-\frac{s \ddot{a}}{8
   a \omega ^4}-\frac{77 i m^3 s^2 \dot{a}}{32 a \omega
   ^7}-\frac{i m^2 s^2 \dot{a}}{2 a \omega ^6}\\&+\frac{11 m^2 \dot{a}
   \dot{s}}{16 a \omega ^6}+\frac{41 i m s^2 \dot{a}}{32 a \omega
   ^5}-\frac{m \dot{a} \dot{s}}{16 a \omega ^5}+\frac{i s^2 \dot{a}}{8 a
   \omega ^4}\\&-\frac{3 \dot{a} \dot{s}}{8 a \omega ^4}-\frac{143 m^4 s
   \dot{a}^2}{64 a^2 \omega ^8}+\frac{103 m^2 s
   \dot{a}^2}{64 a^2 \omega ^6}-\frac{s \dot{a}^2}{8 a^2 \omega ^4}\\&+\frac{15
   m^2 s^3}{16 \omega ^6}+\frac{m s^3}{4 \omega ^5}+\frac{7
   i m s \dot{s}}{8 \omega ^5}-\frac{3 s^3}{16 \omega
   ^4}-\frac{\ddot{s}}{8 \omega ^4}+\frac{i s \dot{s}}{8 \omega ^4}
\end{align*}
\begin{align*}
\phi^{(4)} =&+\frac{\ddddot{a} m}{32 a \omega ^6}-\frac{29 \dot{a} \dddot{a} m^3}{64 a^2 \omega ^8}+\frac{\dot{a}\dddot{a} m^2}{64 a^2 \omega ^7}+\frac{7\dot{a} \dddot{a} m}{32 a^2 \omega ^6}\\
&-\frac{41 \ddot{a}^2 m^3}{128 a^2 \omega^8} -\frac{\ddot{a}^2 m^2}{128 a^2\omega ^7}+\frac{\ddot{a}^2 m}{8 a^2 \omega ^6}\\
&+\frac{951 \dot{a}^2 \ddot{a}m^5}{256 a^3 \omega ^{10}}-\frac{7 \dot{a}^2 \ddot{a} m^4}{64 a^3\omega ^9}-\frac{767 \dot{a}^2\ddot{a} m^3}{256 a^3 \omega ^8}\\
&+\frac{3\dot{a}^2 \ddot{a} m^2}{64 a^3 \omega ^7}+\frac{11\dot{a}^2 \ddot{a} m}{32 a^3 \omega ^6}-\frac{9635 \dot{a}^4 m^7}{2048 a^4 \omega^{12}}\\
&+\frac{10859 \dot{a}^4 m^5}{2048 a^4\omega ^{10}}-\frac{309 \dot{a}^4m^4}{2048 a^4 \omega ^9}+\frac{421 \dot{a}^4 m^6}{2048 a^4 \omega^{11}} \\
&-\frac{337 \dot{a}^4 m^3}{256 a^4 \omega^8}+\frac{\dot{a}^4 m^2}{128a^4 \omega ^7}+\frac{\dot{a}^4 m}{32 a^4 \omega ^6}\\
&+\frac{83 s^4   m}{128 \omega ^6}-\frac{2451 i s \dot{a}^3 m^6}{256 a^3\omega ^{11}}+\frac{75 i s \dot{a}^3 m^5}{256 a^3\omega ^{10}}\\
&+\frac{2431 s^2 \dot{a}^2 m^5}{256a^2 \omega ^{10}}+\frac{2757 i s \dot{a}^3m^4}{256 a^3 \omega ^9}+\frac{143 s^2 \dot{a}^2 m^4}{256 a^2 \omega ^9}\\
&+\frac{385 i s^3 \dot{a} m^4}{64 a\omega ^9}+\frac{387 i \dot{a}^2 \dot{s} m^4}{128 a^2\omega ^9}+\frac{285 i s \dot{a} \ddot{a} m^4}{64 a^2 \omega^9}\\
&-\frac{51 i s \dot{a}^3 m^3}{256 a^3 \omega^8}-\frac{2483 s^2 \dot{a}^2 m^3}{256 a^2 \omega^8}+\frac{95 i s^3 \dot{a} m^3}{64 a \omega ^8}\\
&-\frac{9 i \dot{a}^2 \dot{s} m^3}{32 a^2 \omega ^8}-\frac{143 s\dot{a} \dot{s} m^3}{32 a \omega ^8}+\frac{3 i s \dot{a} \ddot{a} m^3}{32a^2 \omega ^8}\\
&-\frac{117 s^2 \ddot{a} m^3}{64 a \omega^8}-\frac{195s^4 m^3}{128 \omega ^8}-\frac{337 i s \dot{a}^3 m^2}{128 a^3 \omega ^7}\\
&-\frac{103 s^2 \dot{a}^2 m^2}{256a^2 \omega ^7}-\frac{299 i s^3 \dot{a} m^2}{64 a \omega^7}-\frac{367 i \dot{a}^2 \dot{s} m^2}{128 a^2 \omega^7}\\
&-\frac{77 i s^2 \dot{s} m^2}{32 \omega ^7}-\frac{203 is \dot{a} \ddot{a} m^2}{64 a^2 \omega ^7}-\frac{19 i \dot{s} \ddot{a} m^2}{32a \omega ^7}\\
&-\frac{5 s^2 \ddot{a} m^2}{16 a \omega^7}-\frac{19 i \dot{a} \ddot{s} m^2}{32 a \omega ^7}-\frac{11 i s \dddot{a} m^2}{32 a \omega ^7}
\end{align*}
\begin{align*}
\phantom{\phi^{(4)} =}
&-\frac{59 s^4 m^2}{128 \omega^7}+\frac{is \dot{a}^3 m}{32 a^3 \omega ^6}+\frac{121 s^2\dot{a}^2 m}{64 a^2 \omega ^6}+\frac{11\dot{s}^2 m}{32 \omega^6}\\
&-\frac{47i s^3 \dot{a} m}{64 a \omega ^6}+\frac{i \dot{a}^2 \dot{s}m}{8 a^2 \omega ^6}+\frac{99 s \dot{a} \dot{s} m}{32 a \omega^6}-\frac{i s^2 \dot{s} m}{2 \omega ^6}\\
&-\frac{i s \dot{a}\ddot{a} m}{16 a^2 \omega ^6}+\frac{i \dot{s} \ddot{a} m}{32 a \omega^6}+\frac{53 s^2 \ddot{a} m}{64 a \omega ^6}+\frac{i \dot{a} \ddot{s} m}{32 a \omega ^6}\\
&+\frac{9 s \ddot{s} m}{16 \omega ^6}-\frac{i s \dddot{a} m}{32 a\omega ^6}+\frac{i s \dot{a}^3}{16 a^3\omega ^5}+\frac{s^2 \dot{a}^2}{32 a^2 \omega^5}\\
&-\frac{\dot{s}^2}{32 \omega ^5}+\frac{13 i s^3\dot{a}}{32 a \omega ^5}+\frac{7 i \dot{a}^2 \dot{s}}{16a^2 \omega ^5}+\frac{s \dot{a} \dot{s}}{8 a \omega ^5}\\
&+\frac{13 is^2 \dot{s}}{32 \omega ^5}+\frac{i s \dot{a} \ddot{a}}{4 a^2 \omega^5}+\frac{i \dot{s} \ddot{a}}{4 a \omega ^5}+\frac{s^2 \ddot{a}}{16 a\omega ^5}\\
&+\frac{3 i \dot{a} \ddot{s}}{8 a \omega ^5}+\frac{s\ddot{s}}{16 \omega ^5}+\frac{i s \dddot{a}}{16 a \omega^5}+\frac{i \dddot{s}}{16 \omega ^5}+\frac{11 s^4}{128\omega ^5}\,.
\end{align*}
The asymptotic expansion for $h^{II}$ can be easily obtained from that of $h^I$ by performing the exchange $m\mapsto-m$ and $s\mapsto-s$.

It is important to keep in mind that it is not straightforward to compare our asymptotic expansion for $\bm{h}$ with the expressions used in \cite{DFNT}. This is so because of the inherent arbitrariness of the procedure used there. We will discuss the origin of that arbitrariness in section \ref{comparison}. In any case the physical results obtained by both methods match exactly. We will illustrate   this by evaluating the conformal anomaly.

 To  compute the conformal anomaly $C_f$ in the adiabatic renormalization/regularization method, we have to start with a massive field and take the massless limit at the end of the calculation. Therefore,
\be C_f= g^{\mu\nu} \langle T_{\mu\nu}^m\rangle_{\rm ren}  -g_Y\Phi\langle \bar \psi \psi  \rangle_{\rm ren} = -\lim_{m\to 0} m \langle \bar \psi \psi\rangle^{(4)} \ . \ee
Since the renormalized expectation values $\langle T_{\mu\nu} \rangle_{\rm ren}$ require subtractions up to and including the fourth adiabatic order, the adiabatic subtractions for  $\langle \bar \psi \psi\rangle$ above should also include them. Only the fourth order subtraction term  $\langle \bar \psi \psi\rangle^{(4)}$ survives in the massless limit
and produces the anomaly
\be C_f
 =\!-\!\lim_{m \to 0} \frac{m}{\pi^2a^3}
 \!\int^{\infty}_{0}\!\!\!\!\!\mathrm{d}k\, k^2 \!\left( |h^{II}|^2\!-\!|h^{I}|^2 \right)^{(4)} \ ,
\ee
where the subscript $^{(4)}$ is used to indicate that we take only the terms that are of adiabatic order four.
Applying the adiabatic expansion computed in this section and computing the integrals we obtain exactly the same result as in \cite{DFNT}
\bea
 \label{Cf}C_f=&&\frac{\ddddot{a}}{80\pi^2
a}+\frac{s^2\ddot{a}}{8\pi^2a}+\frac{\ddot{a}^2}{80\pi^2a}+\frac{3s\dot{s}\dot{a}}{4\pi^2a}+\frac{s^2\dot{a}^2}{8\pi^2a^2}\\ &&+\frac{3\dot{a}\dddot{a}}{80\pi^2a^2}-\frac{\dot{a}^2\ddot{a}}{60\pi^2a^3}+\frac{s\ddot{s}}{4\pi^2}+\frac{\dot{s}^2}{8\pi^2}+\frac{s^4}{8\pi^2}
\ . \ \ \ \ \ \nonumber
\eea
This result can be nicely rewritten in  covariant form as
\bea \label{Cf2}C_f &=&\frac{1}{2880\pi^2}\left[-11\left(R_{\alpha \beta}R^{\alpha \beta}-\frac13 R^2\right)+6\Box
R\right]
\label{traceanomaly}  \\
&+&\frac{g_Y^2}{8\pi^2}\left[\nabla^{\mu}\Phi\nabla_{\mu}\Phi+2\Phi\Box \Phi +\frac16 \Phi^2 R + g_Y^2\Phi^4\right]. \nonumber
\eea
We finally remark that the covariant expression obtained here for the conformal anomaly shows the covariance of the adiabatic renormalization method. However, intermediate calculations involving the expansion of the modes, like the above expressions for $\phi^{(n)}$, need not to be rephrased in terms of curvature tensors. The same applies for large-$k$ expansion of the modes characterizing the allowed physical states, like the constants $C$, $D$, in Section \ref{scalars}, or the constants $\alpha, \beta$, etc in Section \ref{comparison}. This should not be regarded as a drawback, since generically the renormalized expectation values of local operators are consistent with covariance.

\subsection{Electromagnetic fields}

In this case, in view of \eqref{system}, the functions $\alpha(t)$ and $\beta(t)$ are
\begin{align}\label{equations_Yukawa}
  \alpha(t) & =-\frac{1}{a(t)}(k+q A(t))\,, \\
  \beta(t) & =-m\,,
\end{align}
where $A(t)$ describes a homogeneous electric field ($E= -\dot A$) and $q$  denotes the electric charge. Now the adiabatic order of $A$ is one (see the discussion in \cite{FN}).

In order to compute  the subtraction terms for relevant observables, i.e., the electric current $\langle j^{\mu}\rangle=-q\langle \bar{\psi}\gamma^{\mu}\psi\rangle$, the axial current  $\langle j^{\mu}_A\rangle=\langle \bar{\psi}\gamma^\mu \gamma^{5}\psi\rangle$ and the the energy momentum tensor $\langle T_{\mu\nu} \rangle $, we only need to approximate the solution to adiabatic order two. Therefore,  we need to compute $\bm{h}(t|2)$ by using \eqref{sol_approx}. As in the previous case there is an irrelevant phase factor that we will not give here.  After a straightforward computation we find now
\[
h^I\sim\sqrt{\frac{\omega-k/a}{2\omega}}\Big(1+\phi^{(1)}+\phi^{(2)} \Big)\exp\left(-i\int \omega_2\right)
\]
where $\omega^2=m^2+k^2/a^2$ and
\begin{align*}
  \phi^{(1)}&=-\frac{A q}{2 a \omega }-\frac{A k q}{2 a^2 \omega ^2}+\frac{i m^2 \dot{a}}{4 a \omega^3}-\frac{i \dot{a}}{4 a \omega }-\frac{i k \dot{a}}{4 a^2 \omega^2}\\
  \phi^{(2)}&=\frac{A^2 k q^2}{2 a^3 \omega ^3}-\frac{5 A^2 m^2 q^2}{8 a^2 \omega ^4}+\frac{A^2 q^2}{2 a^2 \omega ^2}-\frac{7 i A k m^2 q \dot{a}}{8 a^3 \omega^5}\\
   &+\frac{i A k q \dot{a}}{2 a^3 \omega ^3}-\frac{3 i A m^2 q\dot{a}}{4 a^2 \omega ^4}+\frac{i A q \dot{a}}{2 a^2 \omega^2}+\frac{i q \dot{A}}{4 a \omega ^2}+\frac{i k q\dot{A}}{4 a^2 \omega ^3}\\
   &-\frac{k \ddot{a}}{8 a^2 \omega ^3}+\frac{m^2 \ddot{a}}{8 a \omega^4}-\frac{\ddot{a}}{8 a \omega ^2}+\frac{3 k m^2\dot{a}^2}{8 a^3 \omega ^5}-\frac{k\dot{a}^2}{8 a^3 \omega ^3}\\
   &-\frac{11 m^4\dot{a}^2}{32 a^2 \omega ^6}+\frac{15 m^2\dot{a}^2}{32 a^2 \omega^4}-\frac{\dot{a}^2}{8 a^2 \omega ^2}
\end{align*}
The asymptotic expansion for $h^{II}$ can be obtained from that of $h^I$ by substituting $a(t)$ for $-a(t)$ and introducing a global minus sign \cite{FN}.

With the help of these expressions one can immediately work out the chiral and conformal anomalies  in the massless limit.
The formal expression for $\langle \nabla_{\mu} j_A^{\mu}\rangle$ has divergences up to second adiabatic order. Therefore, one can write
\be
\langle \nabla_{\mu} j_A^{\mu}\rangle_{\rm ren}=-\lim_{m\to 0} 2 i m \langle \bar{\psi}\gamma^5 \psi\rangle^{(2)} \ . \ee
By expressing  $\langle \nabla_{\mu} j_A^{\mu}\rangle$  in terms of $\bm{h}_k$ we arrive at
\bea
\langle \nabla_{\mu}j_A^{\mu}\rangle_{\rm ren}
&&=-\frac{2 i m}{2 \pi a} \int_{-\infty}^{+\infty} \mathrm{d}k \big( h^{II*} h^{I}- h^{I*} h^{II}\big)^{(2)} \ ,
\nonumber \eea
and using the adiabatic series expansion, it leads immediately to  the axial anomaly in two dimensions
\be
\langle \nabla_{\mu}j_A^{\mu}\rangle_{\rm ren}=\frac{q \dot{A}}{a \pi}= -\frac{q}{2\pi}\epsilon^{\mu\nu}F_{\mu\nu}\ \ , \ee
where $\epsilon^{01}= |g|^{-1/2}= a^{-1} $. A similar calculation leads to the conformal anomaly
\bea
 \langle T^{\mu}_{\mu} \rangle_{\rm ren} &&
 =-\lim_{m\to 0} \frac{m}{2\pi a}\int_{-\infty}^{+\infty} \mathrm{d}k\, \left(h^{I*}  h^{II}+ h^{II*}h^{I}\right)^{(2)}\nonumber.
\eea
Using again the adiabatic expansion and integrating the corresponding adiabatic terms we get
\bea
\langle T^{\mu}_{\mu}\rangle_{\rm ren}=-\frac{R}{24 \pi} \ .
\eea

\section{Conclusions and comments}\label{comparison}

Some comments regarding the method that we have developed in the  paper are in order. First of all we would like to point out that, at variance with the approach used in \cite{LNT, DFNT, FN}, there is no arbitrariness in the expressions for $\bm{h}$ at any iterative order. This is due to the fact that we do not need the adiabatic expansion of the integrands in the exponential term appearing in the expansions
\[
\bm{h}\sim(\bm{\phi}^{(0)}+\cdots+\bm{\phi}^{(n)})\exp(-i\int \omega_n)\,.
\]
Remember that in the approach used in \cite{LNT, DFNT, FN} the main idea was to introduce expansions of the type
\[
\bm{h}\sim(\bm{F}^{(0)}+\cdots+\bm{F}^{(n)})\exp(-i\int \omega^{(0)}+\omega^{(1)}+\cdots+\omega^{(n)})\,,
\]
find the conditions imposed by the equations \eqref{normalization}, \eqref{eqhI}, and \eqref{IIbis} on the unknown terms $\bm{F}^{(j)}$ and $\omega^{(j)}$ at all the adiabatic orders starting from the first and solve for them. The ambiguity that had to be fixed in that context was due to the possibility of adding the time derivative of an arbitrary object inside the integral in the exponent and compensating it by modifying the corresponding $\bm{F}^{(j)}$ (this happens, for instance, if an integration by parts is performed). Although the local subtraction terms required for renormalization were found to be free of any ambiguity, and hence there was no problem to determine ``physical'' quantities  such as $\langle T_{\mu\nu}\rangle_{\rm {ren}}$, the arbitrariness in the adiabatic expansion of the modes is somewhat inconvenient. First because it is not clear a priori that the simplest form of the different terms has been found and, second, because usually the form for the first and second components of $\bm{h}$ do not display some obvious symmetries of the equations \eqref{eqhI} and  \eqref{IIbis}. A third, and more important reason, to rely on the form of the adiabatic expansion that we give in the paper is that it can be used to constrain the {\it allowed physical quantum states} of the Dirac field. As in the case of the massive scalar field \cite{Ivan}, in order to ensure that they lead to finite expectation values for the renormalized stress-energy tensor, they must be of adiabatic order four of higher (in four-dimensional spacetimes).

In the case of a Dirac field coupled to a background scalar through Yukawa interactions, the large-$k$ limit quantum states must obey the asymptotic UV condition
\bea
|h^I|& \sim & \left|\sqrt{\frac{\omega+m}{2\omega}}\Big(1+(\omega-m)\sum_{n=1}^4 \phi^{(n)} \Big)\right| +\mathcal O(k^{-5}) \nonumber \\
&\sim& \frac{1}{\sqrt{2}}+\frac{\alpha(m,s,a)}{2\sqrt{2}\left(k/a\right)}-\frac{\beta(m,s,a)}{8\sqrt{2}\left(k/a\right)^2} \nonumber\\
&-&\frac{ \gamma(m,s,a)}{16 \sqrt{2}\left(k/a\right)^3} +\frac{ \delta(m,s,a)}{128 \sqrt{2}\left(k/a\right)^4}+\mathcal O(k^{-5})\nonumber
\eea
where
\bea
\alpha(m,s,a)&:=&m+s \nonumber\\
\beta(m,s,a)&:=& (m+s)^2\nonumber\\
\gamma(m,s,a)&:=&2 \ddot s+6 \dot s \frac{\dot a}{a}+3(m+s)^3+\frac{1}{3}(m+s) R\nonumber\\
\delta(m,s,a)&:=&(m+s)\Big(8 \ddot s+24 \dot s \frac{\dot a}{a}+11(m+s)^3\nonumber \\ &&\hspace*{3.9cm}+\frac{4}{3}(m+s)R\Big)\nonumber
\ , \eea
and analogously for $|h^{II}|$, after the  the exchange $m\mapsto-m$ and $s\mapsto-s$.

In the case of the electric field background in two spacetime dimensions the vacuum states have to be of adiabatic order two or higher for the renormalized stress-energy tensor to be finite. The corresponding UV condition has the form
\bea |h^I|&\sim & \left|\sqrt{\frac{\omega-k/a}{2\omega}}\Big(1+\phi^{(1)}+\phi^{(2)} \Big)\right| +\mathcal O(k^{-3}) \nonumber \nonumber \\
& \sim& \frac{m}{2\left(k/a\right)}-\frac{mq \left(A/a\right)}{2\left(k/a\right)^2}+\mathcal O(k^{-3}) \nonumber\\
 |h^{II}| &  \sim & 1-\frac{m^2}{8\left(k/a\right)^2}+\mathcal O(k^{-3}) \nonumber
\ . \eea

It is important to point out that the conservation of $\|\bm{h}\|^2$ is a direct and trivial consequence of the dynamics of the Schr\"{o}dinger equation. This is qualitatively different from the situation in the scalar field case that makes use of the WKB approximation. In fact, the attempts to approach the study of the Dirac field in cosmological backgrounds based on the idea of working with second order equations for $\bm{h}$, are hindered by the difficulty of dealing with the conservation of the norm in that context.

From the purely computational point of view it is worth pointing out that in many physical applications (i.e. anomaly computations, etc.) the relevant part of the approximate expression \eqref{sol_approx} is the product of unitary matrices
\[
\bm{U}_0(t)\bm{U}_1(t)\cdots\bm{U}_n(t)\,.
\]
If one wants an adiabatic expansion of this expression it is necessary to get it for the different $\bm{U}_j$. Owing to the square roots appearing in their explicit expressions, in particular, in the off-diagonal terms, (see section \ref{method}) in order to reach a certain adiabatic order $n$ it is usually necessary to compute the adiabatic expansions of the $\omega_j$ to order $2n$. In practice this is not a problem from the computational viewpoint, at least for the examples that we have presented in the paper.

\section*{Acknowledgments}

We thank I. Agullo, S. Pla and F. Torrenti for useful discussions. This work has been supported by the Spanish MINECO research grants FIS2014-57387-C3-3-P; FIS2014-57387-C3-1-P; FIS2017-84440-C2-2-P; FIS2017-84440-C2-1-P, the COST action CA15117 (CANTATA), supported by COST (European Cooperation in Science and Technology), and the Severo Ochoa Program SEV-2014-0398.    A. F. is supported by the Severo Ochoa Ph.D. fellowship SEV-2014-0398-16-1 and the European Social Fund. Some of the computations have been done with the help of Mathematica\texttrademark.


\begin{thebibliography}{99}

\bibitem{parker-toms}L.~Parker and D.~J.~Toms, {\it Quantum Field Theory in Curved Spacetime: Quantized Fields
and Gravity}, Cambridge University Press, Cambridge, England (2009).
\bibitem{fulling} S.~Fulling, {\it Aspects of Quantum Field Theory in Curved Space-Time}, Cambridge University Press, Cambridge, England (1989).
\bibitem{birrell-davies} N.~D.~Birrell  and P.~C.~W.~Davies, {\it Quantum Fields in Curved Space}, Cambridge University Press, Cambridge, England (1982).
\bibitem{Hollands-WaldPR15} S. Hollands and R. M. Wald, {\it Phys. Rept.} {\bf 574}, 1 (2015).
\bibitem{parker66} L.~Parker, {\it The creation of particles in an expanding universe}, Ph.D. thesis, Harvard University (1966).
\bibitem{parker-fulling} L.~Parker and S.~A.~Fulling, {\it Phys.~Rev.~D} {\bf 9}, 341 (1974); S. A. Fulling and L. Parker, {\it Ann.~Phys.} (N.Y.) {\bf 87}, 176 (1974); S. A. Fulling, L. Parker and B. L. Hu, {\it Phys. Rev. D} {\bf 10}, 3905 (1974).  See also, L. Parker, {\it J.~Phys.~A} {\bf 45}, 374023 (2012).
\bibitem{Bunch80} T.~S.~Bunch, {\it J.~Phys.~A} {\bf13}, 1297 (1980).
\bibitem{Birrell} N.~D.~Birrell, {\it Proc.~R.~Soc. B} {\bf 361}, 513 (1978).
\bibitem {Anderson-Parker} P.~R.~Anderson and L.~Parker, {\it Phys.~Rev.~D} {\bf 36}, 2963 (1987).
\bibitem{rio1} A.~del Rio and  J.~Navarro-Salas, {\it Phys.~Rev.~D} {\bf 91}, 064031 (2015).
\bibitem{Anderson} P. R. Anderson, {\it Phys.~Rev.~D} {\bf 32}, 1302 (1985); {\it Phys.~Rev.~D} {\bf 33}, 1567 (1986); P.~R.~Anderson and W.~Eaker, {\it Phys. Rev. D} {\bf 61}, 024003 (1999); S.~Habib, C.~Molina-Paris and E.~Mottola, {\it Phys.~Rev.~D} {\bf 61}, 024010 (1999); J.~D.~Bates and P.~R.~Anderson,  {\it Phys.~Rev.~D} {\bf 82}, 024018 (2010).
\bibitem{Kluger91} Y. Kluger, J. M. Eisenberg, B. Svetitsky, F. Cooper and E. Mottola, {\it Phys. Rev. Lett.} {\bf 67}, 2427 (1991).
\bibitem{Anderson08} P. R. Anderson, C. Molina-Par\'{\i}s, D. Evanich and G. B. Cook, {\it Phys. Rev. D} {\bf 78}, 083514 (2008).
\bibitem{LNT} A.~Landete, J.~Navarro-Salas and F.~Torrenti, {\it Phys~Rev.~D} {\bf 88}, 061501(R) (2013); {\it Phys.~Rev.~D} {\bf 89}, 044030 (2014).
\bibitem{RNT} A.~del Rio, J.~Navarro-Salas and F.~Torrenti, {\it Phys.~Rev.~D} {\bf 90}, 084017 (2014).
\bibitem{Ghosh:2016} S.~Ghosh, {\it Phys.~Rev.~D} {\bf 91}, 124075 (2015); {\it Phys.~Rev.~D} {\bf 93},   044032 (2016).
\bibitem{DFNT} A.~del Rio, A. Ferreiro, J.~Navarro-Salas and F.~Torrenti, {\it Phys.~Rev.~D} {\bf 95}, 105003 (2017).
\bibitem{FN} A. Ferreiro and J. Navarro-Salas,  {\it Phys. Rev. D} {\bf 97}, 125012 (2018).
\bibitem{parker68} L.~Parker, {\it Phys.~Rev.~Lett.} {\bf 21}, 562 (1968); {\it Phys.~Rev.~D} {\bf 183}, 1057 (1969); {\it Phys.~Rev.~D} {\bf 3}, 346 (1971);{\it J.~Phys.~Conf.~Ser.} {\bf  600} no.1, 012001 (2015).
\bibitem{Hawking75} S. W. Hawking, {\it Commun. Math. Phys. } {\bf 43}, 199 (1975).
\bibitem{Waldbook} R.~M.~Wald, {\it Quantum Field Theory in Curved Space-time and Black Hole Thermodynamics}, University of Chicago Press, Chicago, (1994).
\bibitem{fabbri-navarro} A. Fabbri and J. Navarro-Salas, {\it Modeling black hole evaporation}, Imperial College Press, London (2005).
\bibitem{BD79} T. S. Bunch and P. C. W. Davies, {\it Proc. R. Soc. Lond. A} {\bf 360}, 117 (1978).
\bibitem{Anderson2} P. R. Anderson, C. Molina-Paris and E. Mottola, {\it Phys. Rev. D} {\bf 72}, 043515 (2005).
\bibitem{Ivan} I. Agullo, W. Nelson and A. Ashtekar {\it Phys. Rev. D} {\bf 91},  064051 (2015).
\bibitem{Hollands-WaldCMP}  S. Hollands and R. M. Wald, {\it Commun. Math. Phys.} {\bf 223}, 289 (2001); {\it Commun. Math. Phys.} {\bf 231}, 309 (2002), {\it Rev. Math. Phys.} {\bf 17}, 227 (2005).
\bibitem{Pirk93}  K. Pirk, {\it Phys. Rev. D} {\bf  48}, 3779 (1993).
\bibitem{Hollands01} S. Hollands, {\it Commun. Math. Phys.} {\bf 216}, 635 (2001).
\bibitem{Fewster-Verch13} C. J. Fewster and R. Verch, {\it Class. Quant. Grav.} {\bf 30}, 235027 (2013).
\end{thebibliography}
\end{document}